\documentclass[10pt,journal,doublecolumn]{IEEEtran}
\usepackage[para]{threeparttable}
\usepackage{stfloats}
\usepackage{cite}
\usepackage{amsmath,amssymb,placeins}
\usepackage{algorithmic}
\usepackage{graphicx}
\usepackage{textcomp}
\usepackage[mathscr]{euscript}
\usepackage{lipsum}
\usepackage{cuted}
\usepackage{mathtools}
\usepackage{multicol}
\usepackage{array}
\usepackage{makecell}
\usepackage{diagbox}
\usepackage[symbol]{footmisc}
\usepackage{caption}
\usepackage{tabularx}

\usepackage[mathscr]{euscript}
\usepackage[usenames]{color}
\usepackage{amsfonts}
\usepackage{latexsym}
\usepackage{subfigure}
\usepackage[format=hang]{subfig}
\usepackage{caption}
\usepackage{floatrow}
\usepackage{multirow}
\usepackage{epstopdf}
\newtheorem{theorem}{{{\textit{Theorem}}}}

\newtheorem{lemma}{{{\textit{Lemma}}}}
\newtheorem{corollary}{{{{\textit{Corollary}}}}}

\newtheorem{definition}{{{\textit{Definition}}}}

\newtheorem{remark}{{{\textit{Remark}}}}

\newtheorem{example}{{{\textit{Example}}}}

\def\BibTeX{{\rm B\kern-.05em{\sc i\kern-.025em b}\kern-.08em
		T\kern-.1667em\lower.7ex\hbox{E}\kern-.125emX}}
		
		\def\@fnsymbol#1{\ensuremath{\ifcase#1\or *\or \dagger\or \ddagger\or
   \mathsection\or \mathparagraph\or \|\or **\or \dagger\dagger
   \or \ddagger\ddagger \else\@ctrerr\fi}}

\begin{document}
\title{A Direct {\color{black}and New} Construction of Near-Optimal Multiple ZCZ Sequence Sets}
\author{Nishant Kumar, Sudhan Majhi, \IEEEmembership{Senior Member, IEEE}, and Ashish K. Upadhyay}

\IEEEpeerreviewmaketitle
\maketitle
\begin{abstract}
In this paper, for the first time, we present a direct {\color{black} and new} construction of multiple zero-correlation zone (ZCZ) sequence sets with inter-set zero-cross correlation zone (ZCCZ) from generalised Boolean function. {\color{black} Tang \emph{et al.} in their 2010 paper, proposed an open problem to construct $N$ binary ZCZ sequence sets such that each of these ZCZ sequence sets is optimal and if the union of these $N$ sets is taken then that union is again an optimal ZCZ sequence set. The proposed construction partially settles this open problem by presenting a construction of optimal ZCZ sequence sets such that their union is a near-optimal ZCZ sequence set. Further, the performance parameter of each binary ZCZ sequence set in the proposed construction is $1$ and tends to $1$ for their union.} The proposed construction is presented by a two-layer graphical representation and compared with the existing state-of-the-art.
{\color{black} Finally, novel multi-cluster quasi synchronous-code division multiple access (QS-CDMA) system model is provided by using the proposed multiple ZCZ sequence sets.} 
\end{abstract}
%\IEEEpeerreviewmaketitle
\begin{IEEEkeywords}
Generalised Boolean function (GBF), zero-cross correlation zone (ZCCZ), zero-correlation zone (ZCZ), multiple ZCZ sequence sets, multi-cluster QS-CDMA system.
\end{IEEEkeywords}
%\vspace{-.3cm}
\section{Introduction}\label{sec:intro}
\IEEEPARstart{A}{}{\color{black} set of two sequences is called Golay complementary pair (GCP) if sum of their auto-correlation is zero except at zero shift \cite{Golay,AdhikaryZCP2020,AdhikaryNew2020,AdhikaryZCP2018}. However, the lengths of the binary GCPs are quite limited, and therefore, Z-complementary pairs (ZCPs) were introduced by Fan \emph{et al.} \cite{fan2007}.} %A ZCP refers to a pair of sequences of the same length $N$ having zero aperiodic autocorrelation sums for all time shifts inside a zone, called zero correlation zone (ZCZ) \cite{AdhikaryZCP2020,AdhikaryNew2020,AdhikaryZCP2018}. %When ZCZ is same as the length of sequence then resultant sequence pair reduces to a GCP.
The idea of ZCPs was generalized to Z-complementary code set (ZCCS) by Feng \emph{et al.} in \cite{FengZCCS2008}. A ZCCS refers to a set of $K$ codes, each of which consists of M constituent sequences of identical length $L$, having ideal aperiodic auto- and cross-correlation properties inside the ZCZ width (Z) \cite{Sarkarpseudo2021,gobinda}. When $Z = L$ and $K = M$, the set is called complete complementary code (CCC) \cite{das1,das2,das3}. %ZCCSs and CCCs have found application in multi-carrier code division multiple access (MC-CDMA) systems to eliminate multiple-access interference (MAI) for multiple users over asynchronous environment.
{\color{black}In wireless communication, the term "near-far effect" refers to the difficulty a receiver has in hearing a weaker signal from a distant source due to co-channel interference, adjacent-channel interference, capture effect, distortion,  etc.} To reduce the “near-far effect” and ensure interference-free communication in asynchronous code division multiple access (CDMA) systems, ZCZ sequences were introduced in the late 1990s \cite{FaSu}.
%\IEEEPARstart{Z}{ERO}-correlation zone (ZCZ) sequences were proposed in
%the late 1990s, aiming to eliminate the “near-far effect” and to achieve interference-free performance in asynchronous code-division multiple access (CDMA) systems \cite{FaSu}. % ZCZ sequences have received extensive research attention for the quasi-synchronous CDMA (QS-CDMA) system.
%ZCZ around the origin plays an important role in quasi synchronous CDMA (QS-CDMA) system \cite{LiGuPa,NiSu}.
{\color{black}The ZCZ sequence sets can solve the problem of synchronization and the interference limitation problem by introducing quasi synchronous CDMA (QS-CDMA) systems. A ZCZ sequence set can be applied in QS-CDMA system to remove multi-access interference (MAI) and multi-path interference (MPI) in the same cell \cite{LiGuPa,NiSu}. However, length of the ZCZ sequence increases with users which increases the transceiver complexity by increasing number of chips per bits. Thus, the construction of multiple ZCZ sequence sets having inter-set zero cross-correlation zone (ZCCZ) may play a crucial role in multi-cluster QS-CDMA systems  which increases number of users without much system complexity.}
%with multi-cell. To remove both the intra-cell interference and the inter-cell interference, the multiple ZCZ sequence sets having inter-set ZCCZ can be used.}
%{\color{black} Further, with the increase in the number of connected devices in the internet of things (IoT) \cite{chen2012challenges}
%and smart homes lead to very dense networks (number of users per meter). In
%such scenarios, more orthogonal sequences are required to separate the devices in the code domain, increasing the set size of the ZCZ sequence.
%The need for increased set size leads to an increase in the length of the signature sequences used for a device or user. This further increases the computation complexity of the system and makes it less spectrally efficient.}% to remove both the intra-cell interference and the inter-cell interference.}

%When the received signal delays within ZCZ, ZCZ sequences can be employed to remove or reduce MAI and multipath interference (MPI) in quasi synchronous CDMA (QS-CDMA) systems \cite{LiGuPa,NiSu}. Although the ZCZ spreading sequences prevent co-channel interference within each cell, inter-cell interference across neighbouring cells is unavoidable \cite{tang2005design}.

%To address the aforementioned shortcoming, the idea of multiple ZCZ sequence sets with inter set zero-cross correlation zone (ZCCZ) has recently been proposed \cite{hayashi2011ternary,TaFa,torii2012new,torii2012optimal,torii2013extension,wang2014novel,wang2017asymmetric,wang2018asymmetric}.
{\color{black} A multiple ZCZ sequence set comprises ZCZ sequence sets as its subsets and the cross-correlation function between two arbitrary sequences from different subsets has either ZCCZ or low cross-correlation zone (LCCZ) \cite{TaFa,ZhPa,ZhDz,X.chen2021construction}. Authors in \cite{ZhPa} and \cite{ZhDz} used generalised bent function and perfectly non-linear functions respectively to construct multiple ZCZ sequence sets. But they tend to achieve only multiple ZCZ sequence set with interset LCCZ instead of ZCCZ.} 
{\color{black}In \cite{TaFa}, Tang \emph{et al.} proposed a method for constructing multiple binary ZCZ sequence sets from mutually orthogonal Golay complementary set (MOGCS) with good inter-set cross-correlation property and provided an open problem for constructing $N$ optimal binary ZCZ sequence set such that the ratio of ZCZ to ZCCZ is $1/2$.}
\par Motivated by the open problem given in \cite{TaFa}, in this letter, first time we propose a direct and new construction of near-optimal multiple ZCZ sequence sets with ZCCZ using a generalised Boolean function (GBF). %{\color{black}Since the proposed construction is based on GBFs, therefore, it is suitable for rapid hardware generation.}
{\color{black}A graphical analysis of our proposed construction
has also been provided. By using the proposed sequence, we proposed multi-cluster uplink QS-CDMA system which can serve  more users by using smaller length of code to achieve spectral-efficient MAI-free communication. The performance of proposed system model is also provided.} %The proposed construction generalizes construction given in \cite{LiGuPa} and it is optimal over several constructions of A-ZCZ sequence sets presented in \cite{hayashi2011ternary,hayashi2011generalized,torii2012new,torii2012optimal,torii2013extension,wang2014novel,wang2018asymmetric,wang2017asymmetric}. %because of optimality of proposed multiple ZCZ sequence sets. A detailed comparison of the proposed work with some existing works is presented in Table \ref{t1}.
%\vspace{-.1cm}
\section{Notations and Definitions}
\subsection{Definition and Correlation Functions}
%\vspace{-.2cm}
Let $\mathbf{a}_1=(a_{10},a_{11},\hdots, a_{1(L-1)})$ and $\mathbf{a}_2=(a_{20},a_{21},\hdots,$ $a_{2(L-1)})$ be two complex sequences of length $L$. For an integer $u$, aperiodic cross-correlation function (ACCF) of $\mathbf{a}_1$ and $\mathbf{a}_2 ~ \text{is}$ 
\begin{equation}\label{equ:cross}
\gamma(\mathbf{a}_1, \mathbf{a}_2)(u)=
\begin{cases}
\sum_{i=0}^{L-1-u}a_{1i}a^{*}_{2(i+u)}, & 0 \leq u < L,
\\
\sum_{i=0}^{L+u-1} a_{1(i-u)}a^{*}_{2i}, & -L< u < 0.  %\\
%0, & \text{otherwise}.
\end{cases}
\end{equation}
Moreover, ACCF is termed as aperiodic auto-correlation function (AACF) if $\mathbf{a}_1=\mathbf{a}_2$ and denoted as $\gamma(\mathbf{a}_1)(u)$.
Next, we define periodic cross-correlation function (PCCF) in terms of ACCF as $\phi(\mathbf{a}_1,\mathbf{a}_2)(u)=\gamma(\mathbf{a}_1,\mathbf{a}_2)(u)+\gamma^*(\mathbf{a}_2,\mathbf{a}_1)(L-u).$
%\begin{equation}
%\phi(\mathbf{a}_1,\mathbf{a}_2)(u)=\gamma(\mathbf{a}_1,\mathbf{a}_2)(u)+\gamma^*(\mathbf{a}_2,\mathbf{a}_1)(L-u).
%\end{equation}
 \begin{definition}
Let $\mathbf{C}=\{\mathbf{C}_0,\mathbf{C}_1, \hdots ,\mathbf{C}_{P-1}\}$ be a set of $P$ codes having $M$ rows and $L$ columns. Define
$ \mathbf{C_{\eta}}=[\mathbf{a}_{0}^{\eta}~\mathbf{a}_{1}^{\eta}~\hdots~ \mathbf{a}_{M-1}^{\eta}]^T_{M\times L},$
%\begin{align}
 %   \mathbf{C_{\eta}}=[\mathbf{a}_{0}^{\eta}~\mathbf{a}_{1}^{\eta}~\hdots~ \mathbf{a}_{M-1}^{\eta}]^T_{M\times L}, \label{eq:3}
%\end{align}
where $\mathbf{a}_\nu^\eta$ ($0\leq \nu \leq M-1,0 \leq \eta \leq P-1$) is the $\nu$th row sequence and $[\cdot]^T$ represents transpose of matrix $[\cdot]$. Then ACCF of two codes is defined as %$\mathbf{C_{\eta_1}},\mathbf{C_{\eta_2}}\in \textbf{C}$ 
\begin{equation}
    \gamma(\mathbf{C_{\eta_1}},\mathbf{C_{\eta_2}})(u)=\sum_{\nu=0}^{M-1}{\gamma(\mathbf{a}_{\nu}^{\eta_1}, \mathbf{a}_{\nu}^{\eta_2})(u)}.
\end{equation}
\end{definition}
\vspace{-.2cm}
\begin{definition}
  Let \textbf{C} be a collection of codes satisfying
  %\begin{center} 
   \begin{equation}
       \gamma(\mathbf{C_{\eta_1}},\mathbf{C_{\eta_2}})(u)=
      \begin{cases}
        {LM}, & \eta_1=\eta_2 \ \text{and}\ u=0,\\
        0,  & \eta_1=\eta_2 \ \text{and}\ 0<|u|<L, \\
        0,  & \eta_1\neq \eta_2 \ \text{and}\ |u|<L.
      \end{cases}
    \end{equation}  
  %\end{center}
  If $P=M$ then  $\textbf{C}$ is known as $(P,P,L)$-CCC.
\end{definition}

\begin{definition}
  Let ${\boldsymbol{\mathscr{Z}}^l}=\{\mathbf{z}^l_0,\mathbf{z}^l_1,\hdots,\mathbf{z}^l_{K-1}\}$ be a set of $K$ sequences. Then, ${\boldsymbol{\mathscr{Z}}^l}$ is called $(K,Z,L)$-ZCZ sequence set if
  \begin{equation}
      \phi(\mathbf{z}^l_i,\mathbf{z}^l_j)(u)=
      \begin{cases}
       0, & i=j\ \text{and}\ 1\leq |u|{\color{black}\leq} Z,\\
       0, & i\neq j \ \text{and}\ 0\leq |u|{\color{black}\leq} Z,\\
       L, & i=j\ \text{and}\ u=0,
       \end{cases}
  \end{equation}
  where $0\leq i,j\leq K-1$ and $Z$ is termed as ZCZ width.
\end{definition} 
\begin{definition}
Let $\boldsymbol{\mathscr{Z}}=\{\boldsymbol{\mathscr{Z}}^1,\boldsymbol{\mathscr{Z}}^2,\hdots,\boldsymbol{\mathscr{Z}}^N\}$ be a collection of $N$, $(K,Z,L)$-ZCZ sequence sets then $\boldsymbol{\mathscr{Z}}$ is known as a multiple ZCZ sequence set with ZCCZ equal to $Z_c$, if for $0\leq |u| \leq Z_c$,
$\phi(\mathbf{z}^l_i,\mathbf{z}^{l'}_j)(u)=0$, $\forall 1\leq l\neq l' \leq N$.
\end{definition}

{\color{black}
\begin{lemma}[\cite{TaFaMa}]
For a $(K,Z,L)$-ZCZ sequence set $\boldsymbol{\mathscr{Z}}$, the performance parameter $\rho$ satisfies $\rho=\frac{K(Z+1)}{L}\leq 1.$ $\boldsymbol{\mathscr{Z}}$ is said to be optimal if equality holds. It is conjectured that for binary cases $\rho=\frac{2KZ}{L}$. Moreover, if  $\frac{K(Z+2)}{L}=1$(or $\frac{2K(Z+1)}{L}=1$ for binary case) then $\boldsymbol{\mathscr{Z}}$ is said to be near-optimal. \label{def4}
\end{lemma}
}
\vspace{-.4cm}
\subsection{Generalised Boolean Function (GBF) \cite{kumar2022direct}}
We define a complex valued sequence corresponding to a GBF, $f:\{0,1\}^m \longrightarrow\mathbb{Z}_q$ of $m$ variables as $\Psi(f)=\left(\omega^{f_0}, \omega^{f_1}, \hdots, \omega^{f_{2^m-1}}\right),$
%\begin{equation}\label{eqn5}
% \Psi(f)=\left(\omega^{f_0}, \omega^{f_1}, \hdots, \omega^{f_{2^m-1}}\right),
%\end{equation}
where $f_j=f(j_0,j_1,\hdots,j_{m-1})$, $\omega=\exp\left(2\pi\sqrt{-1}/q\right)$, and $(j_0,j_1,\hdots,j_{m-1})$ is the 
binary vector representation of $j$, whereas in the remainder of this letter, $q$ is an even integer not less than $2$. 

\begin{definition}
Let $J = \{j_0, j_1,\hdots, j_{k-1}\} \subset \{0, 1,\hdots , m - 1\}$ and $\mathbf{x}_J = [x_{j_0} , x_{j_1} ,\hdots , x_{j_{k-1}}]$.
For a constant $\mathbf{e} \in \{0, 1\}^k$, $f|_{\mathbf{x}_J=\mathbf{e}}$ is known as restriction of $f$ over $\mathbf{e}$ and is obtained by substituting $x_{j_\beta} = e_\beta$ $(\beta =
0, 1,...,k-1)$ in the function $f$. Moreover, the sequence $\Psi(f|_{\mathbf{x}_J=\mathbf{e}})$ is the same as sequence $\Psi(f)$ of length $2^m$ except for the positions $i_{j_\beta} \neq e_\beta$ for each $0 \leq \beta < k,$ at these positions $\Psi(f|_{\mathbf{x}_J=\mathbf{e}})$ has the zero entries.
\end{definition}
%\vspace{-.2cm}
%\subsection{Quadratic Forms and Graphs }

Let $f$ be GBF of order $r$ over $m$ variables. If $f|_{\mathbf{x}_{J}=\mathbf{e}}$ is a quadratic GBF, then graph of $f|_{\mathbf{x}_{J}=\mathbf{e}}$, i.e., $G(f|_{\mathbf{x}_{J}=\mathbf{e}})$ has vertex set $V = \{x_0, x_1,\hdots,x_{m-1}\}\backslash\{x_{j_0} , x_{j_1} ,$ $\hdots,x_{j_{k-1}}\}$. If there is a term $q_{\beta_1\beta_2}x_{\beta_1}x_{\beta_2}\ (0 \leq \beta_1 < \beta_2 < m,\ x_{\beta_1},
x_{\beta_2} \in V )$ in the GBF $f|_{\mathbf{x}_{J}=\mathbf{e}}$ with $q_{\beta_1\beta_2} \neq 0 \ (q_{\beta_1\beta_2} \in \mathbb{Z}_q)$ then by connecting the vertices $x_{\beta_1}$ and $x_{\beta_2}$ by an edge, the graph $G(f|_{\mathbf{x}_{J}=\mathbf{e}})$ can be obtained. 
For $k = 0,$ the graph of $f|_{\mathbf{x}_{J}=\mathbf{e}}$ is the same as that of $f$.
\vspace{-.4cm}
{\color{black}\subsection{Generalized Reed-Muller Codes}
\begin{definition}
Let $q \geq 2$ and $0 \leq r \leq m,$, then a linear code over $\mathbb{Z}_q$ generated by the $\mathbb{Z}_q$-valued
sequences corresponding to the monomials of degree at most $r$ in
$x_0, x_1, \hdots, x_{m-1}$ is said to be $r$th order generalised Reed-Muller (RM) code and denoted as $RM_q(r;m)$.
\end{definition}
}
\vspace{-.4cm}
\subsection{The Existing Construction of Multiple CCCs}
\begin{lemma}[\cite{Tian2020multiple}]
Let $m, k$, and $s$ are integers with $0 \leq s \leq k \leq m-2$. Define $J_{s}=\left\{j_{k-s}, j_{k-s+1}, \hdots, j_{k-1}\right\}=\{m-s, m-s+1, \hdots, m-1\}$, $J=\left\{j_{0}, j_{1}, \hdots, j_{k-1-s}\right\} \subset \mathbb{Z}_{m-s}$, $I=\left\{i_{0}, i_{1}, \hdots, i_{m-k-1}\right\}=\mathbb{Z}_{m-s} \backslash J$,
$\mathbf{x}=\left[x_{j_{0}}, x_{j_{1}}, \hdots, x_{j_{k-s-1}}\right]$,
$\mathbf{x}_{\mathbf{s}}=\left[x_{j_{k-s}}, x_{j_{k-s+1}}, \hdots, x_{j_{k-1}}\right]$. Let $\pi$ be a permutation on symbols $\{0,1, \hdots, m-k-1\}$.
Let $f$ be a quadratic GBF over the $m$ variables $x_{0}, x_{1}, \hdots, x_{m-1}$, such that for $\mathbf{e} \in\{0,1\}^{k-s}$
\begin{equation}
\left.f\right|_{\mathbf{x}\hspace{-.05cm}=\mathbf{e}}=\frac{q}{2} \hspace{-.05cm}\sum_{\beta=0}^{m-k-2} \hspace{-.25cm}{x_{i_{\pi(\beta)}} x_{i_{\pi(\beta+1)}}}+\hspace{-.25cm}\sum_{\beta=0}^{m-k-1} \hspace{-.25cm}{u_{\beta} x_{i_{\beta}}}+\sum_{\beta=0}^{s-1} \hspace{-.05cm}{v_{\beta} x_{j_{k-s+\beta}}}+v,
\end{equation}
%\begin{equation}
%Q=\frac{q}{2} \sum_{\beta=0}^{m-k-2} x_{i_{\pi(\beta)}} x_{i_{\pi(\beta+1)}},
%\end{equation}
where $u_{\beta} \in \mathbb{Z}_{q}$ $\forall$ $0 \leq \beta \leq m-k-1, v_{\beta} \in \mathbb{Z}_{q}$ $\forall$ $0 \leq \beta \leq s-1$, and $v \in \mathbb{Z}_{q}$
Let $\gamma_1$ and $\gamma_2$ be two end vertices of the path $G(Q)$, $t_{1}=\sum_{\beta=0}^{s-1} b_{k+1+\beta} 2^{\beta}$, $t_{2}=\sum_{\beta=0}^{k} b_{\beta} 2^{\beta}$, where $b_{\beta} \in\{0,1\}$ for $0 \leq \beta \leq k+s$. For the natural order generated by $(t_1,t_2)$, Define the set $S^{\left(t_{1}, t_{2}\right)}$ by
\begin{multline}
\left\{f+\frac{q}{2}\left(\sum_{\beta=0}^{k-1} d_{\beta} x_{j_{\beta}}+d x_{\gamma_1}+\sum_{\beta=0}^{k-1} b_{\beta} x_{j_{\beta}}+\hspace{-.25cm}\sum_{\beta=k-s}^{k-1} \hspace{-.2cm}d_{\beta} b_{s+1+\beta}\right.\right.\\
\left.\left.+b_{k} x_{\gamma_2}\right): d_{\beta}, d \in\{0,1\}\right\}.\label{eq:10}
\end{multline}
Let $\mathbf{S}^{t_{1}}=\left\{S^{\left(t_{1}, t_{2}\right)}: 0 \leq t_{2} \leq 2^{k+1}-1\right\}, 0 \leq t_{1} \leq 2^{s}-1$. Then $\left\{\mathbf{S}^{t_{1}}: 0 \leq t_{1} \leq 2^{s}-1\right\}$ is a collection of $2^{s}$  CCCs, and any two GCSs from different CCCs $\mathbf{S}^{t_{1}}$ and $\mathbf{S}^{t_{1}^{\prime}}$ with $0 \leq$ $t_{1} \neq t_{1}^{\prime} \leq 2^{s}-1$ have a ZCCZ of width $2^{m-s}$.\label{l2}
\end{lemma}
\par For the fixed values of $t_1$ and $t_2$, $S^{(t_{1}, t_{2})}$ is a GCS. Let us denote,
\begin{equation}
S^{(t_{1}, t_{2})}=
    \begin{bmatrix}
      s^{(t_{1}, t_{2})}_0\ s^{(t_{1}, t_{2})}_1\ \hdots\ s^{(t_{1}, t_{2})}_{2^{k+1}-1}
    \end{bmatrix}^T,\label{eq:11}
\end{equation}
where $s^{(t_{1}, t_{2})}_\nu\ (0\leq \nu \leq 2^{k+1}-1)$ is $\nu$th row sequence of $S^{(t_{1}, t_{2})}$.

% *******************Equation in Bottom**************
%\vspace{.5cm}
\begin{figure*}[!b]
\bf{\hrulefill}
 %\begin{align}
%\phi(\Psi(\mathbf{z}^{t_1}_i),\Psi(\mathbf{z}^{t_1}_j))(\tau)=& 2 \cdot \sum_{m=0}^{2^{k+1}-1} \gamma\left(s_{m}^{(t_1,i)}, s_{m}^{(t_1,j)}\right)(\tau)+[(-1)^{h_{l-1}+h_{l}}+(-1)^{h_{2l-1}+h_{0}}] \gamma^{*}\left(s_{0}^{(t_1,j)}, s_{2l-1}^{(t_1,i)}\right)(L-\tau)\nonumber \\
%&+\sum_{m=0}^{2l-2}[(-1)^{h_{m}+h_{m+1}}+(-1)^{h_{m+2l}+h_{m+1+2l}}] \gamma^{*}\left(s_{m+1}^{(t_1,j)}, s_{m}^{(t_1,i)}\right)(L-\tau) .\label{eq:14}
%\end{align}
%\vspace{-.4cm}
\begin{align}
\phi(\Psi(\mathbf{z}^{t_1}_i),\Psi(\mathbf{z}^{t'_1}_j))(\tau)=& 2 \cdot \sum_{m=0}^{l-1} \gamma\left(s_{m}^{(t_1,i)}, s_{m}^{(t'_1,j)}\right)(\tau)+[(-1)^{h_{l-1}+h_{l}}+(-1)^{h_{2l-1}+h_{0}}] \gamma^{*}\left(s_{0}^{(t'_1,j)}, s_{2l-1}^{(t_1,i)}\right)(L-\tau)\nonumber \\
&+\sum_{m=0}^{l-2}[(-1)^{h_{m}+h_{m+1}}+(-1)^{h_{m+l}+h_{m+1+l}}] \gamma^{*}\left(s_{m+1}^{(t'_1,j)}, s_{m}^{(t_1,i)}\right)(L-\tau).\label{eq:15}
\end{align}
%Note: In \eqref{eq:30},\eqref{eq:14}, and \eqref{eq:15} the value of $l=2^k$.\\
\end{figure*}
%\vspace{-.2cm}
% *******************Equation in Bottom**************

\begin{lemma}[\cite{LiGuPa}]
Let $q=2$ and $x_{0}, x_{1}, \hdots, x_{k}, x_{k+1}$ be $k+2$ binary variables. Also, let $h$ be a Boolean function defined on $x_{0}, x_{1}, \hdots, x_{k}, x_{k+1}$ as follow
\begin{equation}
h=\sum_{\beta=1}^{k+1} c_{\beta} x_{\beta} x_{0}+\sum_{1 \leq \mu<\nu \leq k} d_{\mu \nu} x_{\mu} x_{\nu}+\sum_{\alpha=0}^{k+1} e_{\alpha} x_{\alpha}+e^{\prime},
\end{equation}
where $c_{k+1}=1, c_{\beta} \in \mathbb{Z}_{2}$ for $1 \leq \beta \leq k, d_{\mu \nu}, e_{\alpha}, e^{\prime} \in \mathbb{Z}_{2}$. Let $\boldsymbol{h}$ denotes the binary vector corresponding to function $h$, i.e., $\boldsymbol{h}=\left[h_{0}, h_{1}, \hdots, h_{2^{k+2}-1}\right].$
For $0 \leq \tau \leq 2^{k+1}-1$, we have
\begin{equation}
(-1)^{h_{\tau}+h_{\tau+1}}+(-1)^{h_{\tau+2^{k+1}} +h_{\tau+1+2^{k+1}}}=0,\label{eq:18}
\end{equation}
where the operation in the subscripts is done in modulo $2^{k+2}$.\label{l3}
\end{lemma}
\section{Proposed Construction}
%\vspace{-0.4cm}
In this section, we provide a GBF which generates the required multiple ZCZ sequence sets.
\begin{theorem}
Let $x_0,x_1,\hdots,x_{m+k+1}$ are $m+k+2$ binary variables. Define a GBF $f(x_0,x_1,\hdots,x_{m-1})$ on $m$ variables same as in \emph{Lemma} \ref{l2}, i.e., removing $J=\left\{j_{0}, j_{1},\hdots, j_{k-1-s}\right\}$ having $k-s$ vertices from the graph of $f$ results in $s$ isolated vertices in $J_s$ and a path on $m-k$ vertices in $I$. Define another GBF $h(x_m,x_{m+1},\hdots,x_{m+k+1})$ on $k+2$ variables as
%\begin{equation}
    \begin{equation}
      h=\sum_{r=1}^{k+1}{c_r x_{m+r}x_{m}}+\hspace{-.15cm}\sum_{2\leq \mu<\nu\leq t}\hspace{-.35cm}{d_{\mu \nu}x_{m+\mu} x_{m+\nu}}
      +\sum_{\beta=1}^{k+1}{e_\beta x_{m+\beta}}+e',
    \end{equation}
where $c_{k+1}\neq 0,c_r\in\mathbb{Z}_2$, $1\leq r\leq k$, $d_{\mu\nu},e_\beta\in\mathbb{Z}_2$. %Let $\beta$ and $\gamma$ be end vertices of the path $G(Q)$, $t_{1}=\sum_{\alpha=0}^{s-1} b_{k+1+\alpha} 2^{\alpha}, t_{2}=\sum_{\alpha=0}^{k} b_{\alpha} 2^{\alpha}$, where $b_{\alpha} \in\{0,1\}$ for $0 \leq \alpha \leq k+s$.
For a fixed $t_1$, define a set $\boldsymbol{\mathscr{Z}}^{t_{1}}=\{\Psi(\mathbf{z}^{t_{1}}_{t_2}): 0 \leq t_{2} \leq 2^{k+1}-1\}$ by
\begin{multline}
\vspace{-.5cm}
\hspace*{-.3cm}\left\{f+h+\frac{q}{2}\left(\sum_{\beta=0}^{k-1} x_{m+\beta} x_{j_{\beta}}\hspace{-.1cm}+\hspace{-.3cm}\sum_{\beta=k-s}^{k-1} \hspace{-.2cm}x_{m+\beta} b_{s+1+\beta}+\hspace{-.1cm}\sum_{\beta=0}^{k-1} b_{\beta} x_{j_{\beta}}\right.\right.\\
\left.\left.+x_{m+k} x_{\gamma_1}+b_{k}x_{\gamma_2} \right)\right\}.\label{eq:13}
\end{multline}
Then $\boldsymbol{\mathscr{Z}}=\left\{\boldsymbol{\mathscr{Z}}^{t_{1}}\right.$ $\left.: 0 \leq t_{1} \leq 2^{s}-1\right\}$ is a set of $2^{s}~(2^{k+1},2^m$ ,$2^{m+k+2})$-ZCZ sequence sets with ZCCZ equals to $\color{black}2^{m-s}-1$.\label{th1}
\end{theorem}
\begin{IEEEproof}
Using \eqref{eq:10}, \eqref{eq:11}, \eqref{eq:13} and taking natural order generated by $t_2$, we get $\boldsymbol{\mathscr{Z}}^{t_1}=\left[\boldsymbol{\mathscr{Z}_0^{t_1}},\boldsymbol{\mathscr{Z}}^{t_1}_1\right]$, where $\left[\boldsymbol{\mathscr{Z}}^{t_1}_0,\boldsymbol{\mathscr{Z}}^{t_1}_1\right]$ is horizontal concatenation of matrices $\boldsymbol{\mathscr{Z}}^{t_1}_{r-1}$, for $1\leq r\leq 2$ and defined as,
%\begin{align*}
%\begin{center}
  %  &\boldsymbol{\mathscr{Z}}^{t_1}_0=
 %   \begin{bmatrix}
 %     s_{0}^{(t_1,0)}\omega^{h_{0}} & s_{1}^{(t_1,0)}\omega^{h_{1}} & \hdots & s_{l-1}^{(t_1,0)}\omega^{h_{l-1}}\\[.3cm]
   %   s_{0}^{(t_1,1)}\omega^{h_{0}} & s_{1}^{(t_1,1)}\omega^{h_{1}} & \hdots & s_{l-1}^{(t_1,1)}\omega^{h_{l-1}}\\
    %  \vdots & \vdots & \ddots & \vdots\\
    %  s_{0}^{(t_1,l-1)}\omega^{h_{0}} & s_{1}^{(t_1,l-1)}\omega^{h_{1}} & \hdots & s_{l-1}^{(t_1,l-1)}\omega^{h_{l-1}}
   % \end{bmatrix} ,\nonumber
   % \end{align*}
    \begin{align*}
    &\boldsymbol{\mathscr{Z}}^{t_1}_{r-1}=\hspace{-.1cm}
    \begin{bmatrix}
      s_{0}^{(t_1,0)}\omega^{h_{rl}} & \hspace{-.4cm} s_{1}^{(t_1,0)}\omega^{h_{rl+1}} & \hspace{-.4cm}\hdots & \hspace{-.4cm} s_{l-1}^{(t_1,0)}\omega^{h_{(r+1)l-1}}\\[.3cm]
      s_{0}^{(t_1,1)}\omega^{h_{rl}} & \hspace{-.4cm} s_{1}^{(t_1,1)}\omega^{h_{rl+1}} & \hspace{-.4cm}\hdots &\hspace{-.4cm} s_{l-1}^{(t_1,1)}\omega^{h_{(r+1)l-1}}\\
      \vdots &\hspace{-.4cm} \vdots &\hspace{-.4cm} \ddots &\hspace{-.4cm} \vdots\\
      s_{0}^{(t_1,l-1)}\omega^{h_{rl}} &\hspace{-.3cm} s_{1}^{(t_1,l-1)}\omega^{h_{rl+1}} & \hspace{-.4cm}\hdots &\hspace{-.3cm} s_{l-1}^{(t_1,l-1)}\omega^{h_{(r+1)l-1}}\label{eq:30}
    \end{bmatrix},
 % \end{center}
\end{align*}
where $l=2^{k+1}$. Now, we need to prove that $\boldsymbol{\mathscr{Z}}^{t_1}$ is a $(2^{k+1},2^m$ $,2^{m+k+2})$-ZCZ sequence set.
For $0\leq i,j\leq 2^{k+1}-1$, the value of  $\phi(\Psi(\mathbf{z}^{t_1}_i),\Psi(\mathbf{z}^{t_1}_j))(\tau)$ at time shift $0\leq\tau\leq 2^m$ is given by \eqref{eq:15} for $t_1=t'_1$. Next, by \eqref{eq:18}, \eqref{eq:15} for $t_1=t'_1$ and aperiodic sum property of CCCs, we get,
\vspace{-.3cm}
\begin{align*}
    \phi(\Psi(\mathbf{z}^{t_1}_i),\Psi(\mathbf{z}^{t_1}_j))(\tau)&=2 \cdot \sum_{m=0}^{2^{k+1}-1} \gamma\left(s_{m}^{(t_1,i)}, s_{m}^{(t_1,j)}\right)(\tau)\\
    %\vspace{-.3cm}
    &= \begin{cases}
     2^{k+m+2}, & \text{if} \ \tau=0 \ \text{and}\ i=j, \\
      0, & \text{otherwise}.\nonumber
    \end{cases}
\end{align*}
%\vspace{-.2cm}
\begin{table*}[b]
%\tiny
\centering
\caption{Comparison of the proposed construction with \cite{torii2013extension,torii2012optimal,wang2017asymmetric,wang2018asymmetric,X.chen2021construction}.}
\resizebox{\textwidth}{!}{
\begin{threeparttable}
\begin{tabular}{|l|l|l|l|l|l|l|}
\hline
         Ref.
         & Method                   & Parameter\tnote{1} & Optimality\tnote{2} & ZCCZ & No. of sets & Constraints  %&Optimality\tnote{3}             
         \\ \hline
          
         \cite[Th. 1]{torii2013extension}     & Indirect              & $(L,M-1,LP)$       & No& $2M-1$      & $N$ & $N =\lfloor\frac{T}{M}\rfloor > 1,L = KM, M > 1, K > 1$ %& No  
         \\ \hline
         
       \cite[Th. 2]{torii2013extension}         &    Indirect        & $(T,M,TL)$       & No & $TL$      & $N$ & $N =\lfloor\frac{T}{M}\rfloor > 1,L = KM, M > 1, K > 1$ %& No   
       \\ \hline

        \cite{torii2012optimal}             & Indirect             & $(M,M-1,PM)$  & Yes & $PM-1$ &$N$  & $N =\lfloor\frac{T}{M}\rfloor, N > 1, M > 1$ %& No
        \\
       
       \hline
        \cite{wang2017asymmetric}             & Indirect              &$(L,P,TLP)$ & No & $TLP$               &$T$ & $gcd(T,P) = 1, gcd (L,P) = 1 (or L|P or P|L)$ %&No   
        \\
        
        \hline
        \cite{wang2018asymmetric}             & Indirect             & $(2M,Z,2TP)$ & No & $2TP$ &$T$  & $\lfloor\frac{P-2}{Z}\rfloor=M~ \text{or}~\lfloor\frac{P-1}{Z}\rfloor=M,Z\leq2 $ %&No
        \\        \hline
        
        \cite{X.chen2021construction}             & Indirect             & $(N^2,N,N)$ & Yes & $Z+1$ &$M$  & $N$ is order of DFT matrix, $N=M(Z+1)$ %&Yes
        \\        \hline

        This paper & Direct & $(2^{k+1},2^m,2^{m+k+2})$ & Yes & $\color{black}2^{m-s}-1$ & $2^s$ & $0\leq s\leq k\leq m-2$ %&Yes
        \\
        \hline
\end{tabular}
\begin{tablenotes}
\item[1] Parameter of each ZCZ sequence set.\nolinebreak
\item[2] Optimality of each ZCZ sequence set.
%\item[3] Optimality of ZCZ sequence set obtained by union.
\end{tablenotes}
\end{threeparttable}}\label{t1}
\end{table*}
\!\!Which proves that $\boldsymbol{\mathscr{Z}}^{t_1}$ is a $(2^{k+1},2^m,2^{m+k+2})$-ZCZ sequence sets $\forall ~0\leq t_1 \leq 2^s-1$. Now, let $0\leq t_1\neq t'_1 < 2^s$ and $0\leq i,j\leq 2^{k+1}-1$ then for $0\leq\tau\leq \color{black}2^{m-s}-1$, the value of $\phi(\Psi(\mathbf{z}^{t_1}_i),\Psi(\mathbf{z}^{t'_1}_j))(\tau)$ is given by \eqref{eq:15}. Now, by \eqref{eq:18}, \eqref{eq:15} and ZCCZ property of CCCs in \emph{Lemma} \ref{l2}, we get $\phi(\Psi(\mathbf{z}^{t_1}_i),\Psi(\mathbf{z}^{t'_1}_j))(\tau)=0, ~~~\forall~ 0\leq\tau\leq {\color{black}2^{m-s}-1}.$
\end{IEEEproof}
%\begin{equation*}
 %   \phi(\Psi(\mathbf{z}^{t_1}_i),\Psi(\mathbf{z}^{t'_1}_j))(\tau)=0, ~~~\forall~ 0\leq\tau\leq {\color{black}2^{m-s}-1}.
%\end{equation*}

%\begin{remark} \emph{Theorem} \ref{th1} constructed $2^s$ ZCZ sequence sets with parameter $(2^{k+1},2^m,2^{m+k+2})$ having common ZCZ equals to $\color{black}2^{m-s}-1.$ Since ${2^{k+1}\cdot2^m}/2^{m+k+2}=1/2$ and $Z_c={\color{black}2^{m-s}-1=(Z+1)/N}.$
%\end{remark}
\begin{remark}
  Since the set of isolated vertices in \emph{Theorem} \ref{th1} contribute to multipleness of constructed multiple ZCZ sequence set. Hence, for $s=0$, i.e., $J_s=\phi$ in \emph{Theorem} \ref{th1}, our construction reduces to construction in \cite{LiGuPa}. Therefore, construction provided in \cite{LiGuPa} is a special case of the proposed construction.
\end{remark}
\begin{corollary}
Set of all the ZCZ sequences in \emph{Theorem} $\ref{th1}$, i.e., $\{\Psi(\mathbf{z}^{t_{1}}_{t_2}): 0 \leq t_{2} \leq 2^{k+1}-1,~ 0 \leq t_{1} \leq 2^{s}-1\}$ is a {\color{black}near-optimal} $(2^{k+s+1},{\color{black}2^{m-s}-1},2^{m+k+2})$-ZCZ sequence set.
\end{corollary}
%\begin{IEEEproof}
%Directly follows from \emph{Theorem} \ref{th1}.
%\end{IEEEproof}
{\color{black}\begin{remark}
It can easily be calculated that performance parameter for each $\boldsymbol{\mathscr{Z}}^{t_1}$ is $\rho_1=1$ and for their union is $\rho_2\leq1$. Further, $\rho_2$ tends to $1$ for larger value of $m$.
\end{remark}
}
%\begin{remark}
%It is the first time in the literature that the direct construction of optimal multiple ZCZ sequence sets {\color{black}is} provided such that their union is a {\color{black}near-optimal} ZCZ sequence set. Which makes our construction advantageous over several constructions of A-ZCZ sequence sets which are presented in the literature \cite{hayashi2011ternary,hayashi2011generalized,torii2012new,torii2012optimal,torii2013extension,wang2014novel,wang2018asymmetric,wang2017asymmetric}. The detailed comparison of the proposed work is provided in Table \ref{t1}. 
%\end{remark}
\begin{remark}
From equation \eqref{eq:13}, it can be seen that the proposed multiple ZCZ sequence sets are obtained from second order cosets of generalised RM code. Since, RM codes have efficient encoding, good error correction properties and important practical advantage of being easy to decode \cite{DaJe}. Hence, our proposed construction has advantage over any other non-GBF based construction.
\end{remark}
{\color{black}
\begin{remark}
From \eqref{eq:13}, it can be observed that the highest degree of monomials is $2$, therefore the computational complexity of proposed construction for each ZCZ sequence set is of order two, i.e., $O(n^2)$. 
\end{remark}
}
\vspace{-.45cm}
\section{Graphical Approach}
%\vspace{-.2cm}
This section interprets the proposed construction with graphical point of view.

\begin{figure}[ht]
    \centering
    \includegraphics[width=\textwidth,height=1.8in]{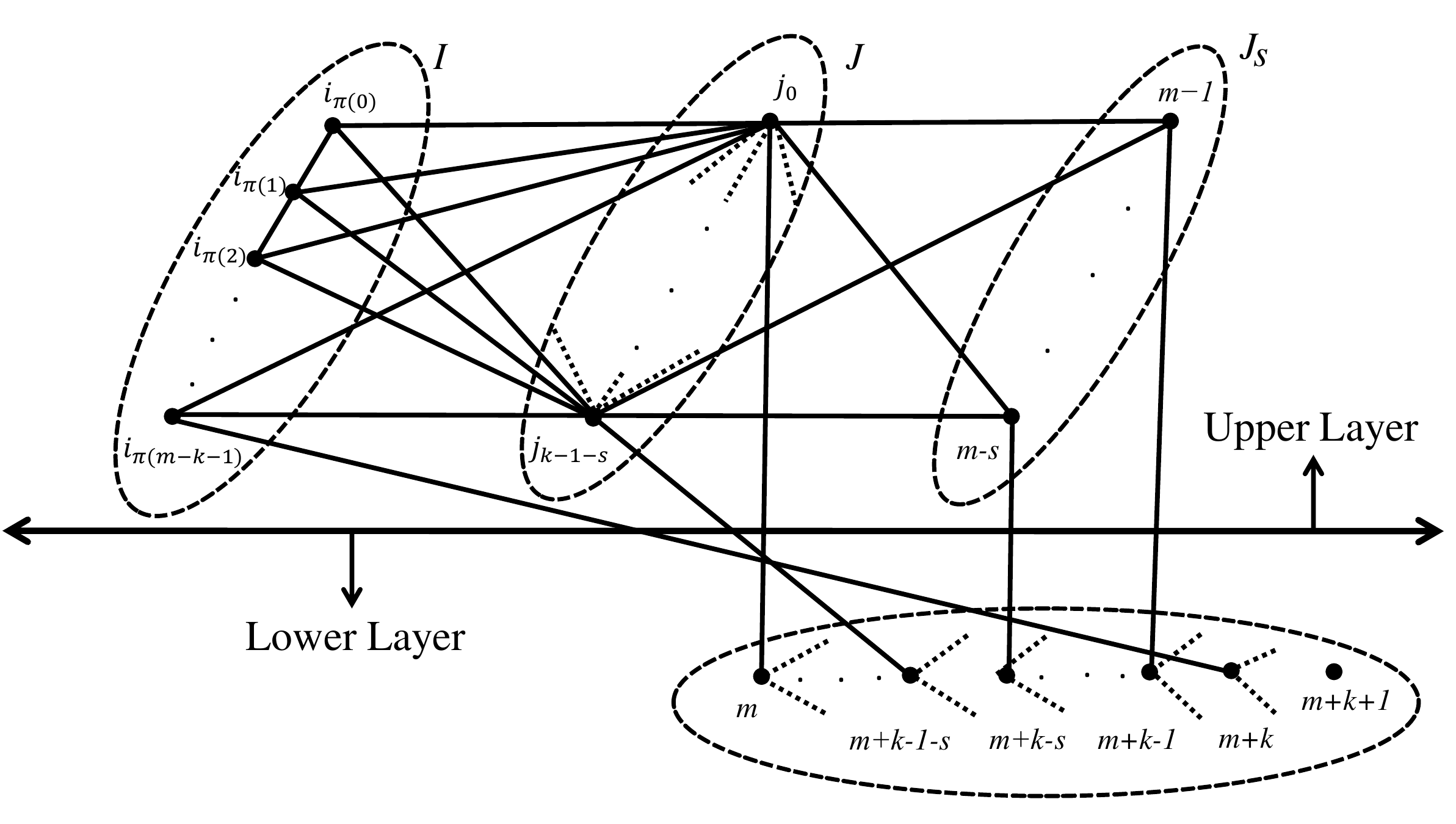}
    \caption{{\color{black}Graphical representation of %\eqref{eq:13}.
    $\{f+h+\frac{q}{2}(\sum_{\alpha=0}^{k-1} x_{m+\alpha} x_{j_{\alpha}}$ $+x_{m+k} x_{\beta})\}.$
    }}
    \label{f1}
\end{figure}

%\vspace{-.5cm}
Fig. \ref{f1} depicts a graphical representation of \eqref{eq:13}. %$\{f+h+\frac{q}{2}(\sum_{\beta=0}^{k-1} x_{m+\beta} x_{j_{\beta}}$ $+x_{m+k} x_{\gamma_1})\}$.
The graph has a two-layered structure with a horizontal straight line which is separating the upper and bottom layers.
The upper layer and lower layer {\color{black}correspond} to graphs of Boolean functions $f$ and $h$ respectively. These layers are interconnected through the set of edges
\vspace{-.3cm}
\begin{multline*}
    \{x_{j_0}x_{m},x_{j_1}x_{m+1},\hdots, x_{j_{k-1-s}}x_{m+k-1-s},x_{m-s}x_{m+k-s},\\
    x_{m-s+1}x_{m+k-s+1},\hdots,x_{m-1}x_{m+k-1}\},
\end{multline*}
%\vspace{-.2cm}
 and the vertex $x_{m+k}$ is connected to any of the end vertices of the path in $I$.
 Interestingly, the ZCZ of each ZCZ sequence set is equals to the power of number of vertices in the upper layer of the graph and ZCCZ of ZCZ sequence sets equals to {\color{black}one less than} the power of number of vertices in the upper layer of graph except isolated vertices. 
 \vspace{-.2cm}
\vspace{-.1cm}
\begin{example}\label{ex1}
Let $m=4$, $q=2$, $s=1$, and $k=2$. Assume $J=\{0\}$, $J_s=\{3\}$, $I=\{1,2\}$ and GBFs $f=x_0x_1+x_0x_2+x_0x_3+x_1x_2+x_1+x_2,~h=x_4x_5+x_4x_6+x_4x_7+x_4.$
%\begin{equation}
   %% f&=x_0x_1+x_0x_2+x_0x_3+x_1x_2+x_1+x_2,\\
%h&=x_4x_5+x_4x_6+x_4x_7+x_4.
 %   \end{aligned}
%\end{equation} 
Generate two sequence sets $\boldsymbol{\mathscr{Z}}^{0}$ and $\boldsymbol{\mathscr{Z}}^{1}$ as
\vspace{-.3cm}
\begin{figure}[ht]
    \centering
    \includegraphics[width=2.3in,height=1.3in]{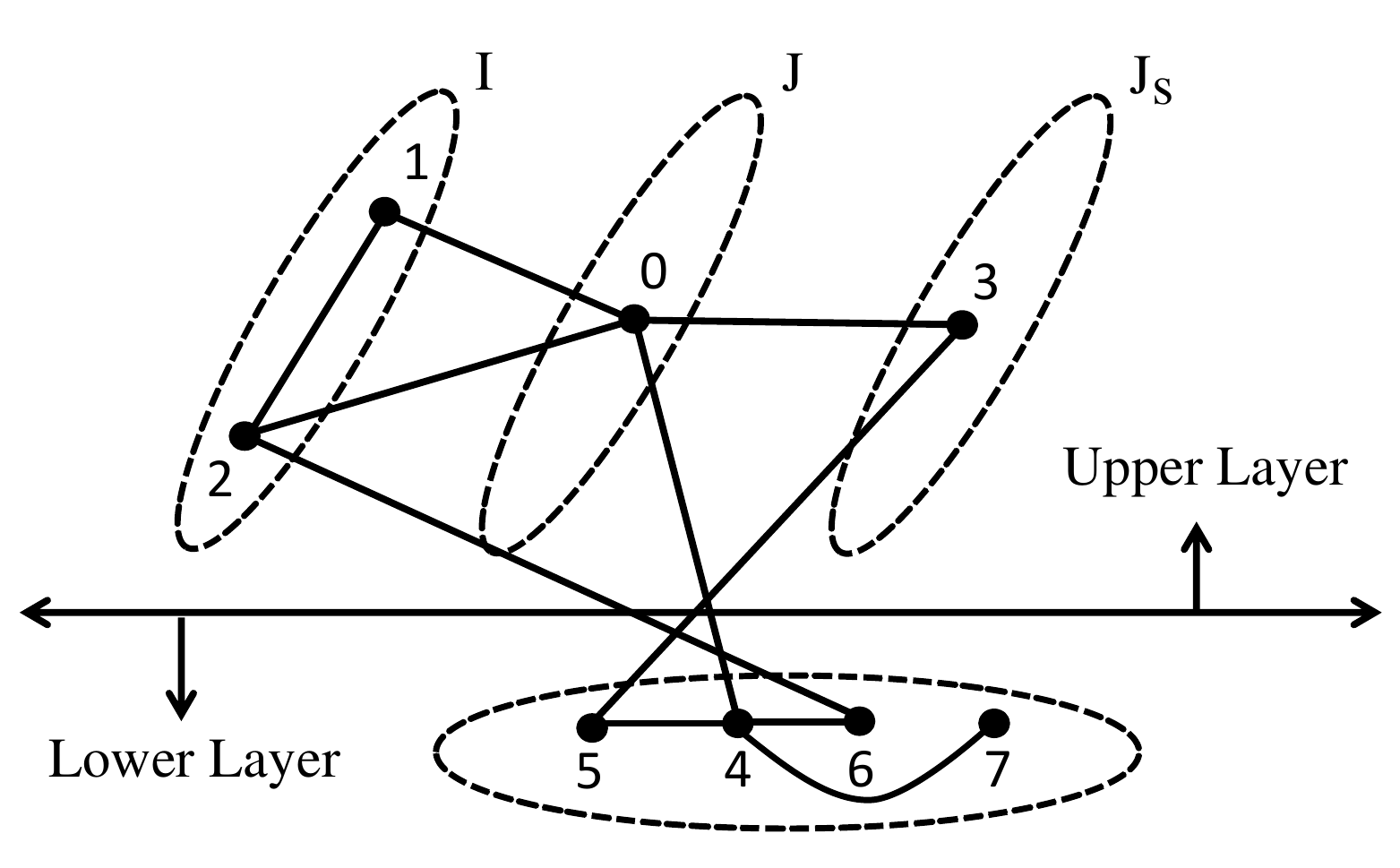}
    \caption{Graphical representation of $f+h+x_0x_4+x_2x_6+x_3x_5.$}
    \label{f2}
\end{figure}
\vspace{-.4cm}
\begin{multline}
  \boldsymbol{\mathscr{Z}}^{0}=\{\Psi(f+h+x_0x_4+x_2x_6+x_3x_5+b_0\cdot x_0+b_1\cdot x_3+b_2\cdot x_1\\
  +0\cdot x_5):b_0,b_1,b_2\in \mathbb{Z}_2\},\\
  %\end{multline*}
  %\begin{multline}
  \boldsymbol{\mathscr{Z}}^{1}=\{\Psi(f+h+x_0x_4+x_2x_6+x_3x_5+b_0\cdot x_0+b_1\cdot x_3+b_2\cdot x_1\\
  +1\cdot x_5):b_0,b_1,b_2\in \mathbb{Z}_2\}.
\end{multline}
Then $\boldsymbol{\mathscr{Z}}^{0}$ and $\boldsymbol{\mathscr{Z}}^{1}$ are two optimal $(8,16,256)$-ZCZ sequence sets having inter-set ZCCZ equals to $7$. {\color{black}Moreover, $\boldsymbol{\mathscr{Z}}=\boldsymbol{\mathscr{Z}}^{0}\cup \boldsymbol{\mathscr{Z}}^{1}$ is a near-optimal $(16,7,256)$-ZCZ sequence set.} In Fig. \ref{f2}, a graph corresponding to quadratic form, i.e., $f+h+x_0x_4+x_2x_6+x_3x_5$ of \emph{Example} \ref{ex1} is presented.
\end{example}
\vspace{-.2cm}
\section{System Model and Performance Analysis}
{\color{black}Consider a dense network scenario as in Fig. \ref{f6} where number of users communicate with a single base station (BS) directly or via an access point (AP). The users communicating with the AP may cause interference to the other AP. All the users communicating nearby AP, are kept in a cluster and each user in that cluster is assigned a signature sequence from the $\mathrm{ZCZ}$ set assigned to that cluster. The total number of clusters that can be served in a cell, depends on the number of multiple $\mathrm{ZCZ}$ sets.
}{\color{black}\!\!\!\!\!\!We present the the system BER for a multi-cluster QS-CDMA system operating on an additive white Gaussian noise (AWGN) channel. We have considered $4$ clusters and $8$ users in each cluster. The multiple ZCZ sequence has 4 sequence sets, one set for each cluster and each set constitutes $8$ ZCZ sequences, which is equal to the number of users in that cluster and the length of each sequence is 256. 
\begin{figure}[ht]
    \centering    \includegraphics[width=6cm,height=2in]{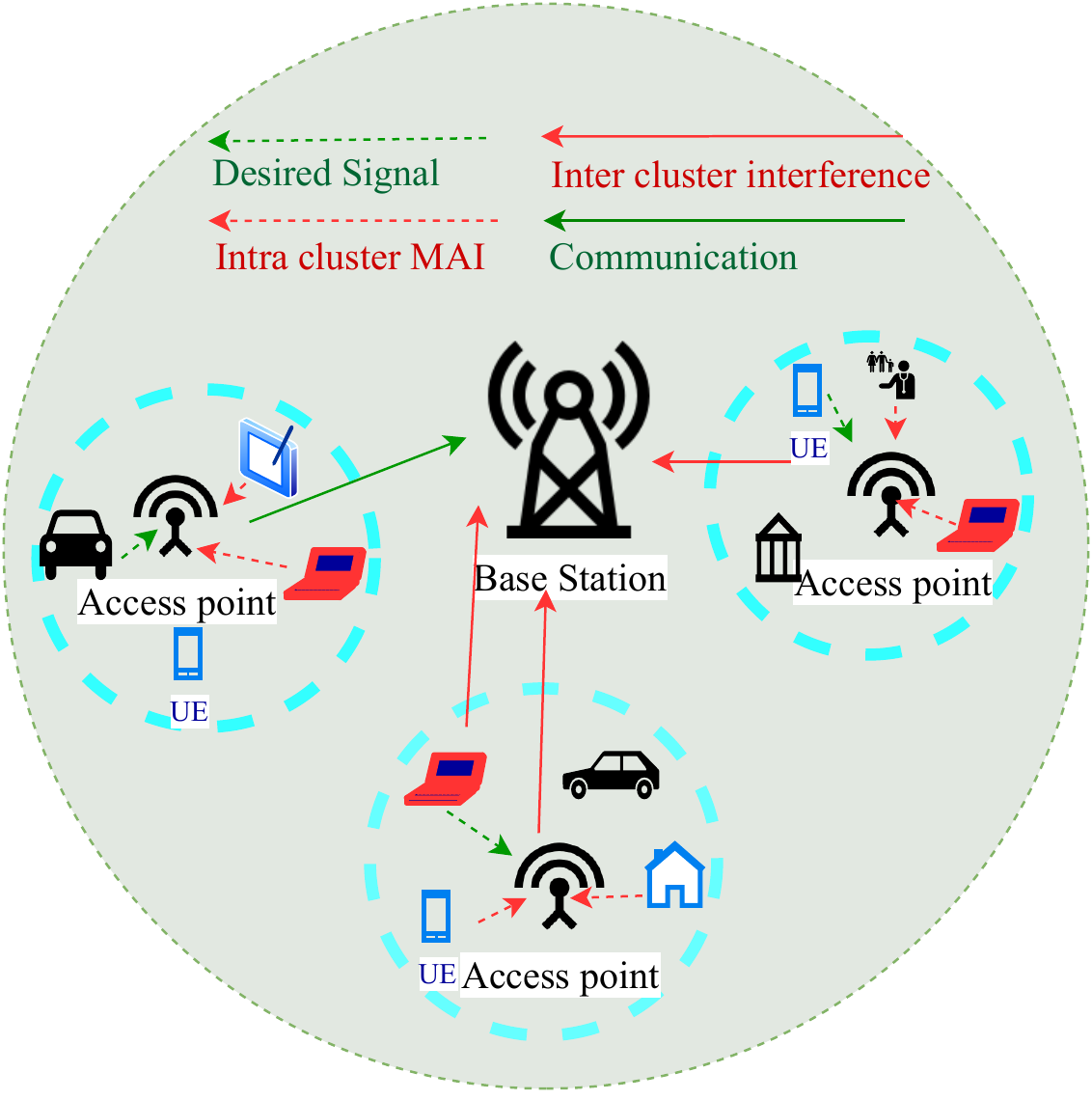}
    \caption{{\color{black} Multi-cluster QS-CDMA system model.}}
    \label{f6}
\end{figure}
The BER performance of one user from each cluster has been provided considering interference from $7$ users and $3$ clusters as depicted in Fig. \ref{f5}. It can be noticed that all the users show almost same BER performance and the BER is as low as $10^{-5}$ for $0 \mathrm{~dB}$ SNR when the simulation was performed for $10^5$ Monte-Carlo iterations with $10^4$ bits. This is because of the processing gain achieved by the system due to spreading by larger sequence length. The delay tolerance of the system is 3 chips. But for accommodating the same number of users in conventional QS-CDMA system using ZCZ, it needs ZCZ sequence of length $2048$. This leads to high processing gain but the spreading efficiency and hence the bandwidth efficiency of the system decreases drastically which makes the system impractical for real world scenario, thus the performance analysis has been omitted.
\vspace{-.2cm}
\begin{figure}[ht]
    \centering    \includegraphics[width=3.4in,height=2in]{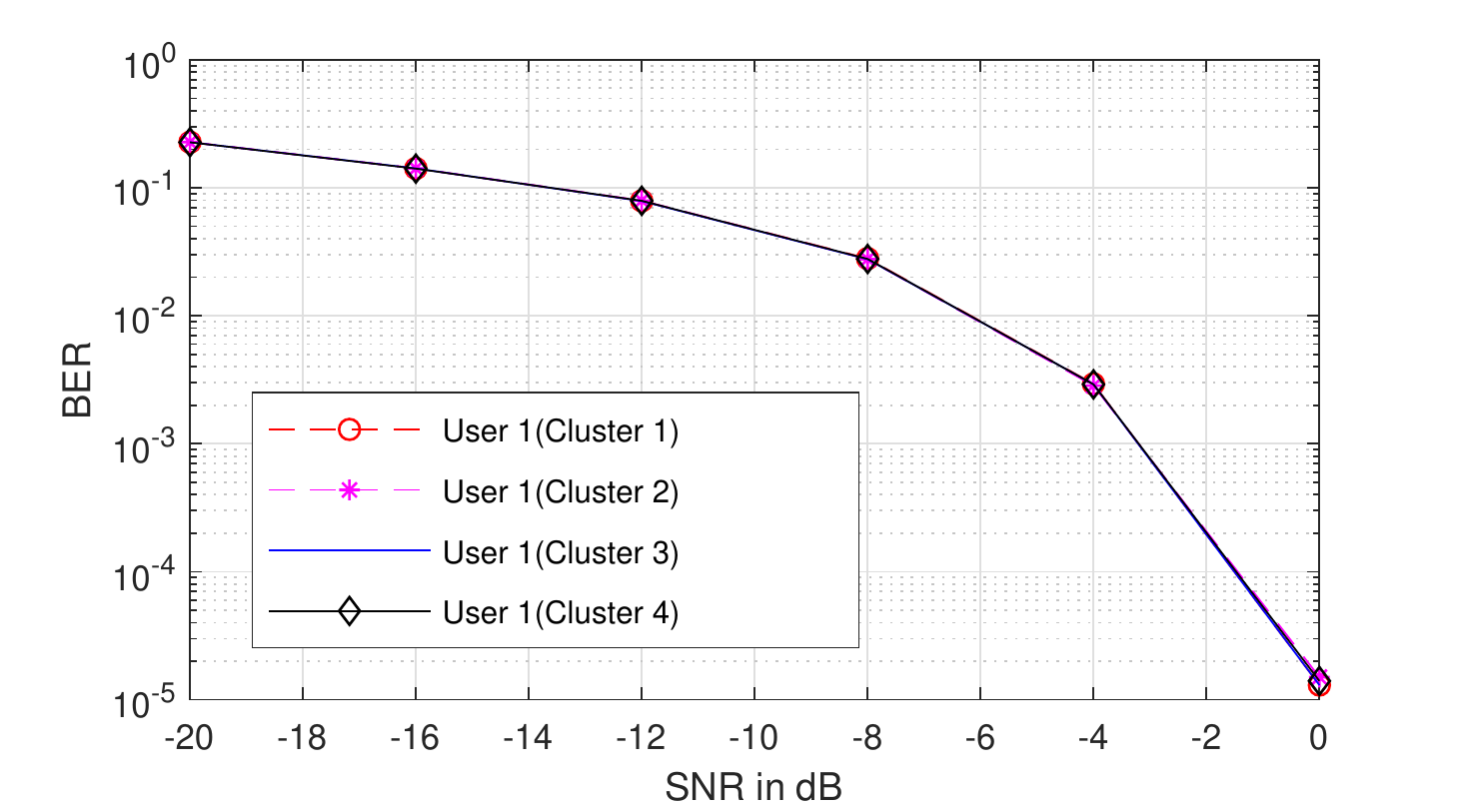}
    \caption{{\color{black}  BER performance of multi-cluster QS-CDMA using multiple ZCZ.}}
    \label{f5}
\end{figure}}
\vspace{-.8cm}
\section{Conclusion}
{\color{black}In this paper, for the first time in the literature, we proposed a direct construction of multiple ZCZ sequence sets having ZCCZ using GBF. {\color{black}A novel multi-cluster uplink QS-CDMA system has also been proposed}. Further, We have covered many lengths which have not been covered before. The computational complexity of the proposed construction for every ZCZ sequence set is $O(n^2)$. Direct construction of multiple ZCZ sequence sets covering some more lengths which have not been covered in the literature will be a great contribution in this direction of research.}

%We further associated a two-layer graph to the proposed set of multiple ZCZ sequence set.
 %The construction in \cite{LiGuPa} appears as a special case of the proposed construction. %Our proposed construction also extends and generalizes the ZCZ construction in \cite{LiGuPa},
 %The proposed construction obtain many new multiple ZCZ sequence sets which have not been discovered before.% A comparison of the proposed construction with existing works is provided in Table \ref{t1}.
\bibliographystyle{IEEEtran}
\bibliography{main}
\end{document}